\def\year{2024}\relax
\documentclass[letterpaper]{article} 
\usepackage{aaai22}  
\usepackage{times}  
\usepackage{helvet}  
\usepackage{courier}  
\usepackage[hyphens]{url}  
\usepackage{graphicx} 
\usepackage{natbib}  
\usepackage{caption} 
\usepackage{amsmath}
\usepackage{mathtools}
\usepackage{url}
\DeclareCaptionStyle{ruled}{labelfont=normalfont,labelsep=colon,strut=off} 
\usepackage{tabularx}
\usepackage{array}
\usepackage{microtype}
\usepackage{lipsum}
\usepackage{enumitem}
\usepackage{booktabs,dcolumn}
\newcolumntype{Z}{>{\centering\let\newline\\\arraybackslash\hspace{0pt}}X}
\usepackage{multirow}
\usepackage{array}
\usepackage{booktabs}
\newcolumntype{C}[1]{>{\centering\let\newline\\\arraybackslash\hspace{0pt}}m{#1}}
\newcolumntype{Z}{>{\centering\let\newline\\\arraybackslash\hspace{0pt}}X}

\usepackage{algorithm}
\usepackage{algorithmic}
\newcommand{\answerYes}[1]{\textcolor{blue}{#1}} 
\newcommand{\answerNo}[1]{\textcolor{teal}{#1}} 
\newcommand{\answerNA}[1]{\textcolor{gray}{#1}} 
 
\usepackage{newfloat}
\usepackage{listings}
\floatstyle{ruled}
\newfloat{listing}{tb}{lst}{}
\floatname{listing}{Listing}

\usepackage{xcolor}
\usepackage{soul}

\title{Exploring Platform Migration Patterns between Twitter and Mastodon: \\ A User Behavior Study}

\author{
    Ujun Jeong\textsuperscript{\rm 1}, Paras Sheth\textsuperscript{\rm 1}, Anique Tahir\textsuperscript{\rm 1}, Faisal Alatawi\textsuperscript{\rm 1}, H. Russell Bernard\textsuperscript{\rm 2}, Huan Liu\textsuperscript{\rm 1}
}
\affiliations{
    \textsuperscript{\rm 1}School of Computing and Augmented Intelligence, Arizona State University\\
    \textsuperscript{\rm 2}Institute for Social Science Research, Arizona State University\\
    \{ujeong1, psheth5, artahir, faalataw, asuruss, huanliu\}@asu.edu 
}

\usepackage{bibentry}
\usepackage{lipsum} 
\usepackage{xcolor} 
\usepackage{fancyhdr}

\definecolor{myred}{RGB}{255,0,0}

\fancypagestyle{titlepagestyle}{
    \fancyhf{} 
    \fancyhead[C]{ 
        \setlength{\fboxsep}{2pt} 
        \fcolorbox{myred}{white}{\textbf{Please cite the ICWSM’24 version of this article}}
    }
}

\begin{document}

\maketitle
\begin{abstract}

A recent surge of users migrating from Twitter to alternative platforms, such as Mastodon, raised questions regarding what migration patterns are, how different platforms impact user behaviors, and how migrated users settle in the migration process. In this study, we elaborate on how we investigate these questions by collecting data over 10,000 users who migrated from Twitter to Mastodon within the first ten weeks following the ownership change of Twitter. Our research is structured in three primary steps. First, we develop algorithms to extract and analyze migration patterns. Second, by leveraging behavioral analysis, we examine the distinct architectures of Twitter and Mastodon to learn how user behaviors correspond with the characteristics of each platform. Last, we determine how particular behavioral factors influence users to stay on Mastodon. We share our findings of user migration, insights, and lessons learned from the user behavior study. 

\end{abstract}

\section{Introduction}

With the proliferation of social media platforms, users tend to be increasingly mobile, moving between different platforms as their needs, preferences, and interests change, a phenomenon known as online platform migration~\cite{zengyan2009cyber}. The ownership change of Twitter on October 27, 2022, created an opportunity to study platform migration as numerous users migrated to Mastodon~\cite{kupferschmidt2022musk, he2023flocking}, a microblogging platform with similar features to Twitter, such as ``toots" and ``boosts" corresponding to ``tweets" and ``retweets," but which operates on a decentralized platform of self-hosted servers~\cite{zignani2018follow}.

\begin{figure}
    \centering
    \includegraphics[width=0.45\textwidth]{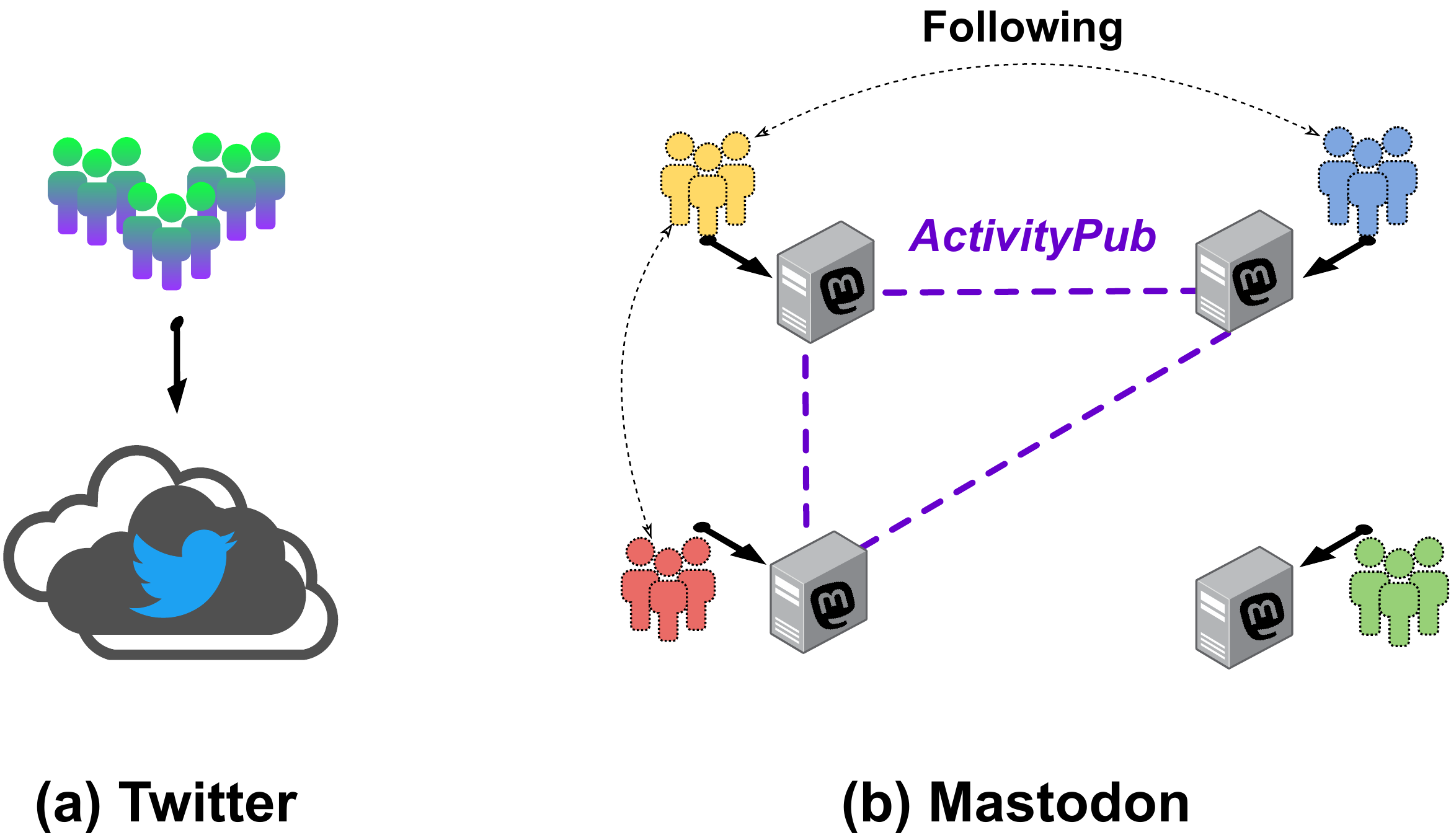}
    \caption{The distinct platform architectures of (a) Twitter, which is a centralized platform, and (b) Mastodon, which employs a decentralized platform with a federated network.}
    \label{fig:mastodon_architecture}
\end{figure}

Despite its decentralized architecture, Mastodon enables users to request follows and share toots, boosts, or favorite requests across servers using the shared Activity Pub protocol~\cite{zignani2018follow}. This facilitates the formation of federated servers and interest-based communities, each managed by distinct moderation policies, as depicted in Figure~\ref{fig:mastodon_architecture}. Mastodon is thus not a replica of Twitter; it provides a unique, community-based user experience and interactions across various servers~\cite{la2021understanding}. Such distinctive features made Mastodon an increasingly popular choice for users seeking a new platform for migration.





Online platform migration has been  researched~\cite{kumar2011understanding, newell2016user, fiesler2020moving, he2023flocking}. Observing the exodus from Twitter, however, we are intrigued to understand user migration patterns: \textbf{(1)} what follows the exodus, i.e., whether migration is persistent or waning - this issue arises because users can maintain accounts on multiple platforms, oscillating their focus between platforms until fully committing to the new platform; \textbf{(2)} whether a platform architecture impacts the behaviors of migrated users - exploring this connection may reveal the interplay between platform differences, subsequent user behaviors post-migration, and varying degrees of user engagement; and \textbf{(3)} what behavioral factors contribute to the sustainability of platform migration is essential for a comprehensive grasp of the dynamics of the migration process.


In this study, we investigate migration patterns by comparing behaviors of users between Twitter and Mastodon, focusing on the dynamics of migrated users switching their attention between the two platforms over time. To facilitate this comparative study, we propose methods to map the accounts of over 10,000 individuals who migrated from Twitter to Mastodon and to determine users' occupational backgrounds. Moreover, we collect data on user behaviors, including user activities and network interactions on each platform during the first ten weeks after Twitter’s leadership change.




\begin{figure}
    \centering
    \includegraphics[width=0.35\textwidth]{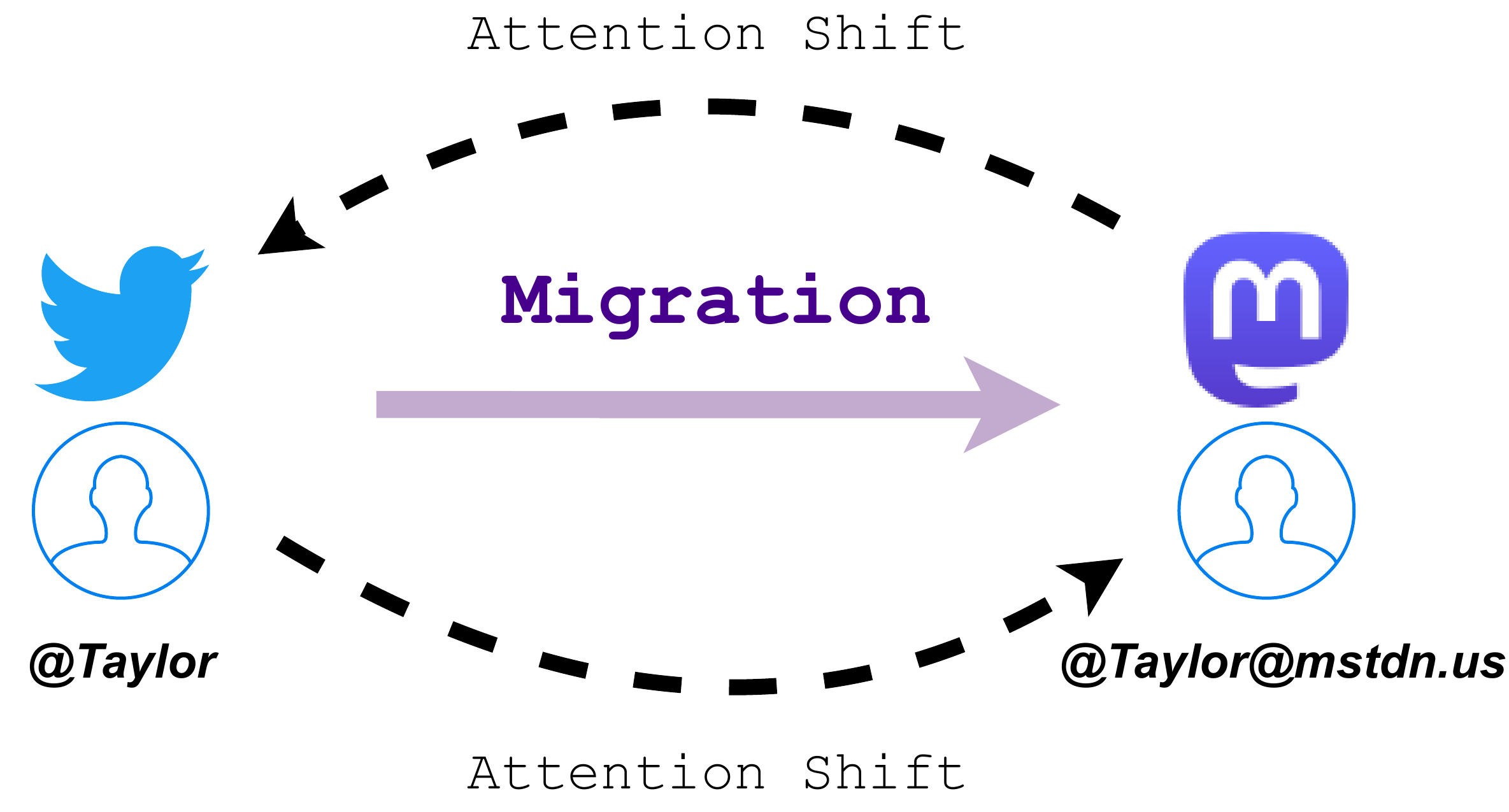}
    \caption{The process of migration, while still maintaining the previous account and shifting attention between platforms.}
    
    \label{twitter_mastodon_user_profile_compare}
\end{figure}
Our study is motivated by three research questions:


\begin{itemize}
    \item \textbf{RQ1}: What are the migration patterns observed between Twitter and Mastodon after Twitter's ownership change?
    \item \textbf{RQ2}: What distinct user behaviors between Twitter and Mastodon arise due to each platform's architecture?
    \item \textbf{RQ3}: What specific behavioral characteristics of migrated users influence their decision to stay on Mastodon?
\end{itemize}  

With respect to \textbf{RQ1}, we examine the evolving migration patterns with users' active status and attention shifts between these platforms over time. Building on this analysis, we correlate migration patterns with notable events, such as Musk's acquisition of Twitter, his promise of stepping down as a CEO, and so on. {Through comparison of discourse and sentiment with migration patterns, we aim to uncover the potential motivations and how they affect migration process. 





In light of \textbf{RQ2}, we compare the behaviors of migrated users on Twitter and Mastodon to understand the disparities between the two platforms. Especially, our experiments include comparing the distribution of users' occupations and hashtags based on their popularity on each platform. This experiment aims to serve as empirical evidence of how the design of each platform affects the engagement of specific user groups more than others.

Regarding \textbf{RQ3}, we focus on the behavioral traits of users who stayed on Mastodon despite notable events drawing attention back to Twitter. By performing statistical analyses, we highlight associations between user retention and various user behaviors, including some distinctive features exclusive to Mastodon. Through this analysis, we shed light on Mastodon's unique appeal to its users and draw insights on certain behavioral factors that promote sustainable migration.

Our main contributions are summarized as follows:
\begin{itemize}
    \item Following the platform policies, we crawl and curate a dataset of over 10,000 users who migrated from Twitter to Mastodon by developing an effective method of mapping accounts between the two platforms.
    \item We propose a novel framework for understanding the intricate dynamics of migration patterns between Twitter and Mastodon, with a focus on the shift of user attention between the two platforms. It enables us to study the interplay among migration factors, the unique platform architectures, and the user behaviors on each platform. 
    \item We present key behavioral factors that promote the sustainability of platform migration to Mastodon, interesting insights for future study of platform migration. 
\end{itemize}


\section{Related Work}

\subsection{Migration Theory and Platform Migration}
Migration has long been a subject of study across social sciences, with the push-pull theory being a key concept in this field~\cite{levitt2007transnational,lee1966theory}. The push-pull theory suggests that factors pushing people away from a location and pulling them toward a new one drive migration decisions. This concept can also be applied to online platform migration~\cite{fiesler2020moving,gerhart2019social, newell2016user}. \citeauthor{newell2016user}~(\citeyear{newell2016user}) studied the motivations through push-pull factors by surveying users who migrated across subreddits during community unrest on Reddit, and found that niche communities play an essential role in attracting users. Factors including platform design, toxicity, moderation policy, presence of friends and community can be motivations for migration~\cite{fiesler2020moving}. Through behavioral analysis, \citeauthor{kumar2011understanding}~(\citeyear{kumar2011understanding}) found significant differences between migrants and random users in three key areas: posting activity, network activity, and Google search ranking. \citeauthor{lHorincz2017collapse}~(\citeyear{lHorincz2017collapse}) found that social capital, such as a high degree of connection and openness of connection, can influence early abandonment of platform. \citeauthor{he2023flocking}~(\citeyear{he2023flocking}) studied the recent surge in users migrating from Twitter to Mastodon. This study presented several findings, including users' tendency to join larger servers, the higher number of imported Twitter followers on these larger servers, and the fact that messages posted by migrated users had relatively lower toxicity on Mastodon than on Twitter.


Mastodon's decentralized architecture attracted many users from traditional centralized platforms like Twitter and Facebook because of its several unique characteristics~\cite{he2023flocking, la2021understanding}. \citeauthor{la2021understanding}~(\citeyear{la2021understanding}) found that connections in Mastodon are  topic-oriented, rather than popularity-driven due to the lack of a recommendation system. Mastodon also emphasizes conversations and interactions over favorites and reshares~\cite{la2022network}. Despite Mastodon's unique appeals over traditional platforms, it faces numerous challenges inherent to decentralized networks.  Information consumers on Mastodon establish most connections and broker information~\cite{la2022information}. The user-driven trend toward the centralization of Mastodon, primarily on a handful of large servers, is also observed~\cite{shaw2020decentralized, zignani2018follow}. Other issues, such as advertising revenue, handling moderation, and the the availability of servers, pose challenges to operating Mastodon~\cite{anaobi2023will, zulli2020rethinking, raman2019challenges}.

Our study diverged from prior platform migration research by examining dynamic migration patterns between Twitter and Mastodon over time. We focused on users' attention shifts between platforms and presented algorithms to infer their migration motivations. Moreover, we analyzed the relation between platform architecture and the behaviors of migrated users, identifying behavioral factors that retain users on Mastodon. This holistic approach deepens our understanding of migration patterns and sustainable platform migration.


\section{Migration Types on Social Media}
\label{migration_definitions}


In social media and migration studies, two types of migration are identified~\cite{hou2020understanding, fiesler2020moving, kumar2011understanding, levitt2007transnational}: (1) \textit{Permanent migration}, where users transition to a new platform, deactivate their original account, and exclusively engage on the new platform; and (2) \textit{Attention migration}, where users maintain presence on both platforms but shift their focus toward one of them.

\paragraph{\textbf{Permanent Migration.}} Let $U_{p_1}$ represent the set of users on the platform $p_1$ and $U_{p_2}$ represent the set of users on the platform $p_2$. A user $u$ is considered to have permanently migrated from platform $p_1$ to platform $p_2$ if two conditions are met: (1) the user $u$ was a member of $p_1$ before time $t$, and (2) the user $u$ is no longer a member of $p_1$ and is a member of $p_2$ at time $t$ (i.e., $u \notin U_{p_1}$ and $u \in U_{p_2}$).


Permanent migration may be a result of profile removal, deletion, or suspension from the original platform.



\paragraph{\textbf{Attention Migration.}} For a user $u$ who is a member of both platforms (i.e., $u \in U_{p_1}$ and $u \in U_{p_2}$) and active on both platforms at time $t_i$, attention migration is said to occur between two distinct times $t_i$ and $t_j$ where $t_j > t_i$ if $u$ becomes inactive on platform $p_1$ while remaining active on platform $p_2$ at time $t_j$. In this case, the user's attention is considered to have migrated from platform $p_1$ to platform $p_2$ during the time interval $\delta = t_j - t_i$.


Attention migration could represent a short-term migration with the potential to evolve into permanent migration over time. To ascertain a user's active status on a platform, we define an active user as:


\paragraph{\textbf{Active User.}} Given a social media platform $p$, a user $u \in U_p$, time $t_j > t_i$, and interval $\delta = t_j - t_i$, $u$ is considered to be active on the platform $p$ at time $t_j$, if the user has performed at least one activity on the site since time $t_i$. Otherwise, the user is considered inactive. User activities include various actions possible on the social media platform, such as posting messages and resharing other users' contents.


Our study recognizes that the permanent migration is not limited to the transition from Twitter to Mastodon, as the reverse direction is also possible. However, we focus on users who initiated migration from Twitter to Mastodon to examine the impact of Twitter's ownership change on migration. For this specific objective, we excluded users who created Mastodon accounts before creating their Twitter accounts.
 
\begin{table}
\centering

\small
\begin{tabular}{cccc}
\toprule
\textbf{Server Name} &  \textbf{\# User}&\textbf{\# Migrant}&\textbf{Domain}\\ 
\midrule
mstdn.social & 176,621& 1,965 &General\\ 
\midrule
mastodon.world  & 146,753& 1,245&General\\ 
\midrule
mas.to  &148,995& 2,180 &General\\ 
\midrule
c.im  &61,854& 397 &General\\
\midrule
masto.ai  &64,908& 452 &Technology\\
\midrule
fosstodon.org  &52,993& 1,230&Technology\\
\midrule
infosec.exchange  &45,532& 2,087 &Technology\\ 
\midrule
sfba.social  &36,972& 325 &General\\ 
\midrule
mindly.social  &33,855& 179 &General\\ 
\midrule
toot.community  &28,227& 273 &General\\
\bottomrule
\end{tabular}
\caption{Top 10 Mastodon servers with most incoming users. We report the number of migrants detected by our method.}
\label{mastodon_server_statistics}
\end{table}

\section{Data Collection}
\label{data_collection}
From October 24, 2022, to January 2, 2023, we identified a total of 10,333 migrated users who have accounts on both Twitter and Mastodon. We removed 56 users (0.5\%) who had Mastodon accounts before Twitter accounts, along with 266 users (2.5\%) whose accounts are inaccessible either on Twitter or Mastodon as of January 2, 2023. This process left  us 10,011 migrated users for our analysis.

\subsection{Mapping Users across Platforms}

Several services, such as Twitodon, aid in merging Twitter and Mastodon accounts for users switching platforms but only support limited account mappings. \citeauthor{he2023flocking} (\citeyear{he2023flocking}) addressed this by analyzing tweets containing Mastodon URLs, confirming matches only when Mastodon and Twitter usernames were identical.  In contrast, our approach allows for the flexible identification of migrated users, even with nonidentical usernames, and incorporates both Mastodon and Twitter as resources for account mapping.


\begin{figure}
    \centering
    \includegraphics[width=0.42\textwidth]{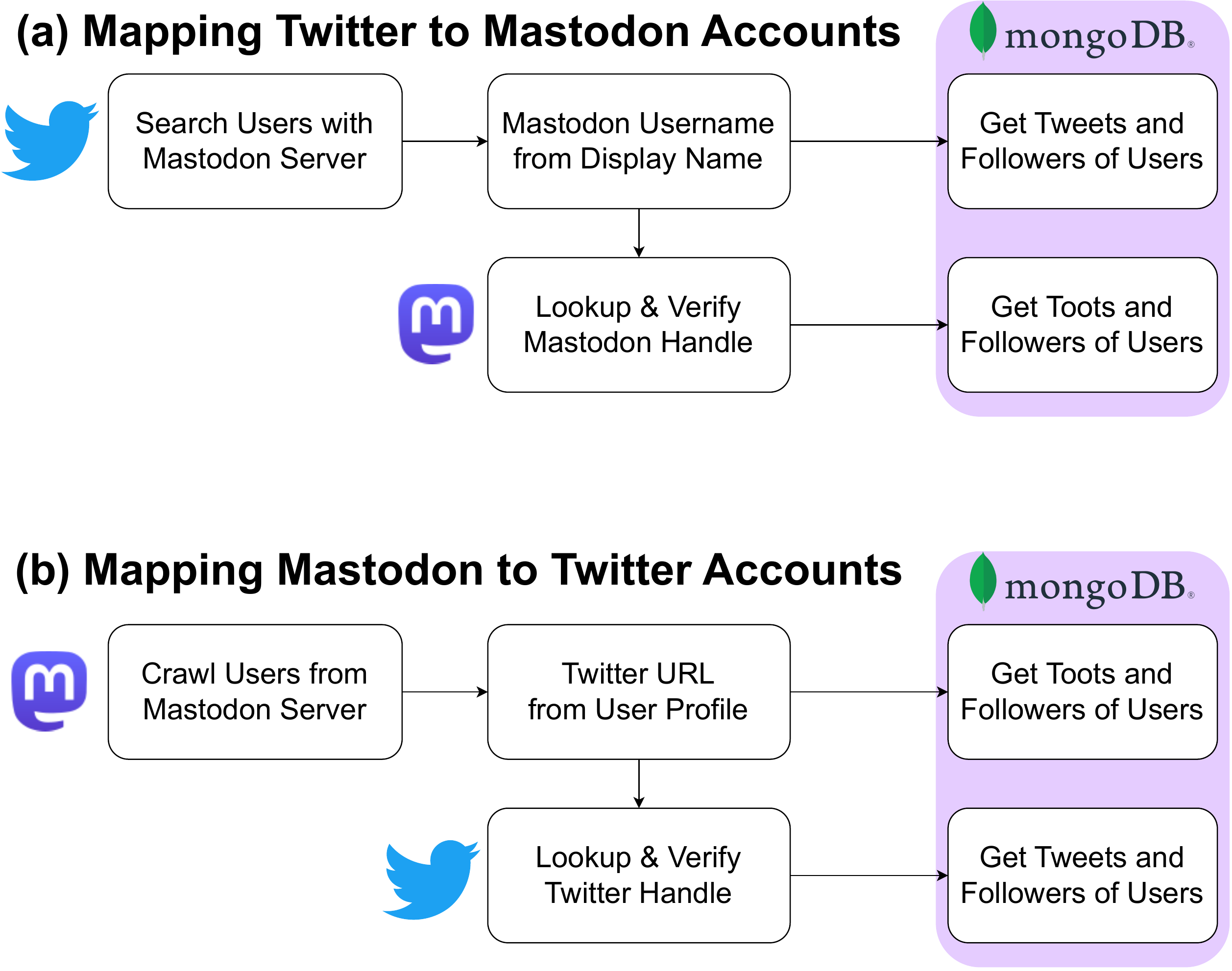}
    \caption{The process of data collection, including mapping accounts, using both Twitter and Mastodon as resources.}
    \label{dataset_collection_pipeline}
\end{figure}

\paragraph{\textbf{Mapping Twitter to Mastodon Accounts.}}
We leverage Twitter users' display names in their profile to map their Mastodon accounts, as these users often disclose their Mastodon usernames and associated Mastodon servers within their display names. We first search for users with the name of any Mastodon server listed in Table~\ref{mastodon_server_statistics}, and we extract the Mastodon usernames from the display names using a regular expression that identifies the word following the @ symbol.

\paragraph{\textbf{Mapping Mastodon to Twitter Accounts.}}
We utilize Mastodon users' profiles, noting that they frequently disclose their additional social media accounts. We first collect user profiles from each Mastodon server. Then, we extract Twitter handles from URLs in user profiles. If a Twitter URL is missing, we search the profile's meta fields for entries marked with Twitter and extract the handle from this value.


It is worth noting that Mastodon's decentralized structure poses challenges in gathering information across all servers~\cite{raman2019challenges}. During our research, several prominent servers, such as \textit{mastodon.social}, temporarily closed new account registrations due to a surge in user traffic. To maintain the integrity of this study, we concentrated on servers that are highly available and accessible when using the Mastodon API. Moreover, we utilized the \textit{instances.social} API to analyze the number of users for each server. Specifically, we narrowed down our server selection to those primarily using English. This allowed us to gather data from the top 10 Mastodon servers that had the most newly registered users during the research period, as depicted in Table~\ref{mastodon_server_statistics}.

\subsection{Collecting User Activities and Network Interactions}

We utilized APIs, with their authenticated credentials, to monitor and collect various behaviors of the migrated users included in our study. The collected data were securely stored in a MongoDB database in an anonymized form, employing field-level encryption as a security measure.

Our data collection encompassed a variety of activities from detected migrants such as tweets, retweets, toots, and boosts. Our data collection efforts included gathering the messages posted during the research period on the public timelines of both Twitter and Mastodon. This process yielded 1.1M tweets and 0.5M retweets from Twitter, along with around 0.6M toots and 0.4M boosts from Mastodon. 

We expanded our data collection to encompass followers of detected migrants on Twitter and Mastodon, resulting in obtaining nearly all available profiles of the followers of the migrants on these platforms: 85.42\% of the total 5.3M followers on Twitter and 88.72\% of the total 2.9M followers on Mastodon. We also collected data on interactions among the migrants on Mastodon, such as replies to the postings, and the profiles of users who participated in these replies.


\subsection{Collecting and Grouping User Occupations}

Since users may not always include occupations in their profiles, we utilized Stanford CoreNLP~\cite{manning2014stanford}, a pretrained model capable of identifying occupation titles through named entity recognition. We first filtered users by examining Twitter profiles for occupation titles, and if these are unavailable, we referred to their corresponding Mastodon profiles. Then, we employed the UK's Standard Occupational Classification (SOC 2010) system to assign an occupation per user, widely adopted in prior research on social media~\cite{sloan2015tweets}. We assigned the user's occupation title to the first corresponding three-digit SOC code from the system's dictionary. If no match was found, we applied the method suggested by \citeauthor{turrell2019transforming}~(\citeyear{turrell2019transforming}) utilizing the user's profile bio. This method identified the SOC codes for 5,817 of 10,011 users (58.1\%). To manage numerous SOC codes, we focused on the first digit (e.g., 1xx), which denotes the nine major groups. Two annotators classified these major groups for a random 300-user subset, resulting in a Cohen's Kappa of 77.82\%. Our approach achieved F1 score of 65.26\%, effectively predicting nine major groups. Figure~\ref{occupation_distribution} displays the distribution of occupational groups among migrated users, with 2xx (Professional Occupations) and 3xx (Associate Professionals and Technicians) groups being the most prominent.


\begin{figure*}
    \centering
    \includegraphics[width=0.87\textwidth]{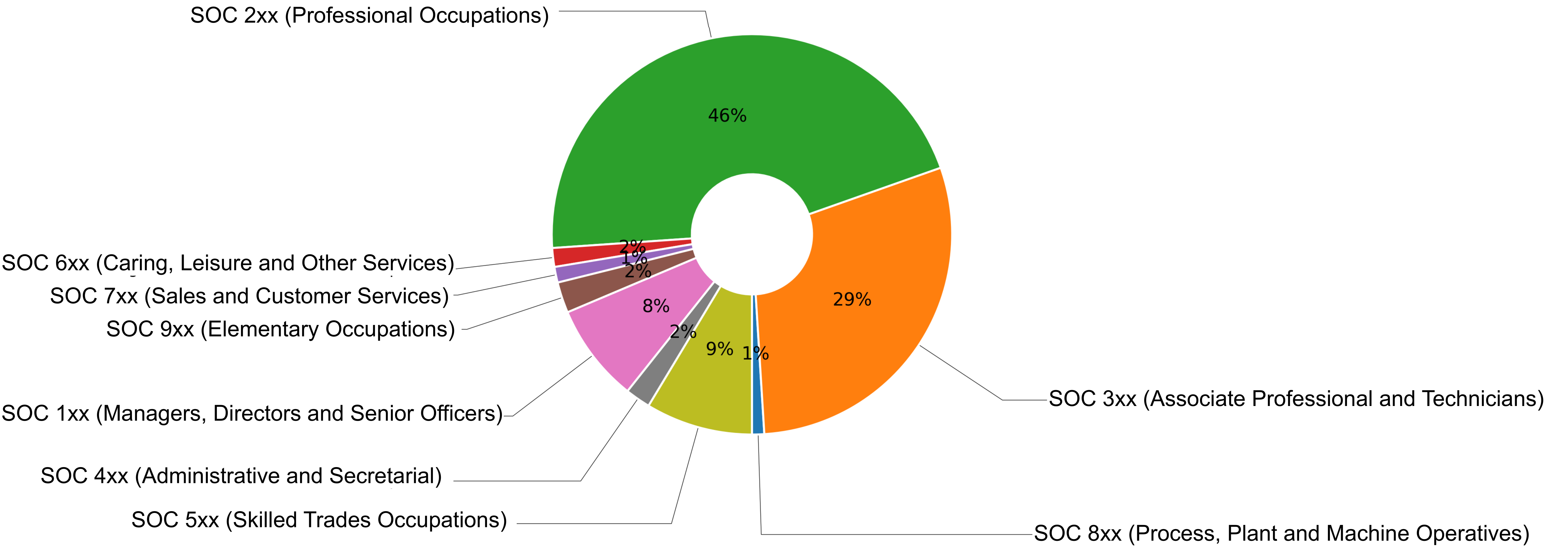}
´    \caption{The pie chart illustrates the distribution of the distribution of nine major groups, each accompanied by corresponding tags. The major groups are based on the first digit of the UK's Standard Occupational Classification (SOC 2010) code, which has been assigned to each user. Further descriptions regarding nine major groups in the UK's SOC 2010 are provided in Appendix.}
    \label{occupation_distribution}
\end{figure*}

\section{RQ1: What Are the Migration Patterns between Twitter and Mastodon?}

\begin{figure*}
    \centering
    \includegraphics[width=0.889\textwidth]{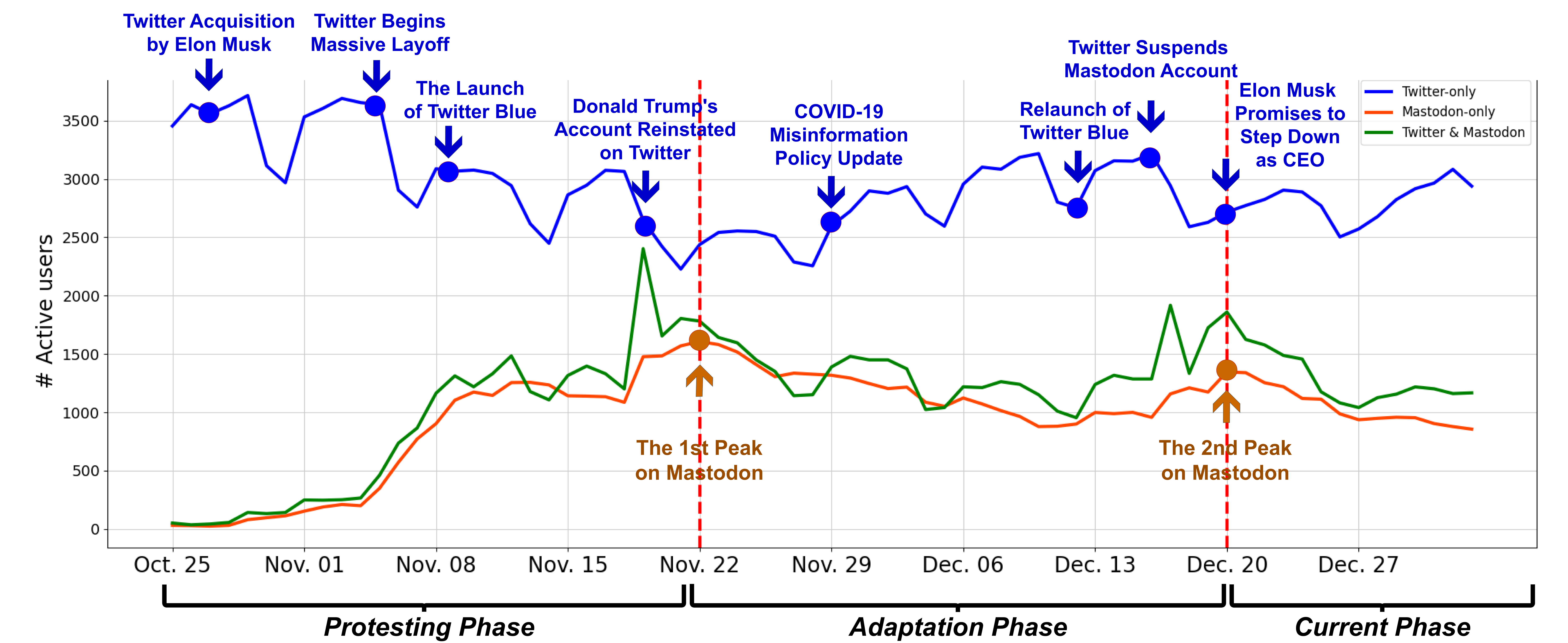}
    
    \caption{Trends in daily active users tagged with major events. Twitter-only (blue) and Mastodon-only (orange) lines indicate the number of users active exclusively on one platform, while Twitter \& Mastodon (green) represents users active on both platforms. The $x$-axis denotes a particular date $t_j$, when we assess whether a user was active within the interval of $\delta$ = \textit{1 day}. Red dashed lines highlight the key moments, where the temporal shifts overlap among the three trends (blue, orange, and green).}
    \label{mastodon_active_user_trend}
\end{figure*}

\begin{figure*}
    \centering
    \includegraphics[width=0.87\textwidth]{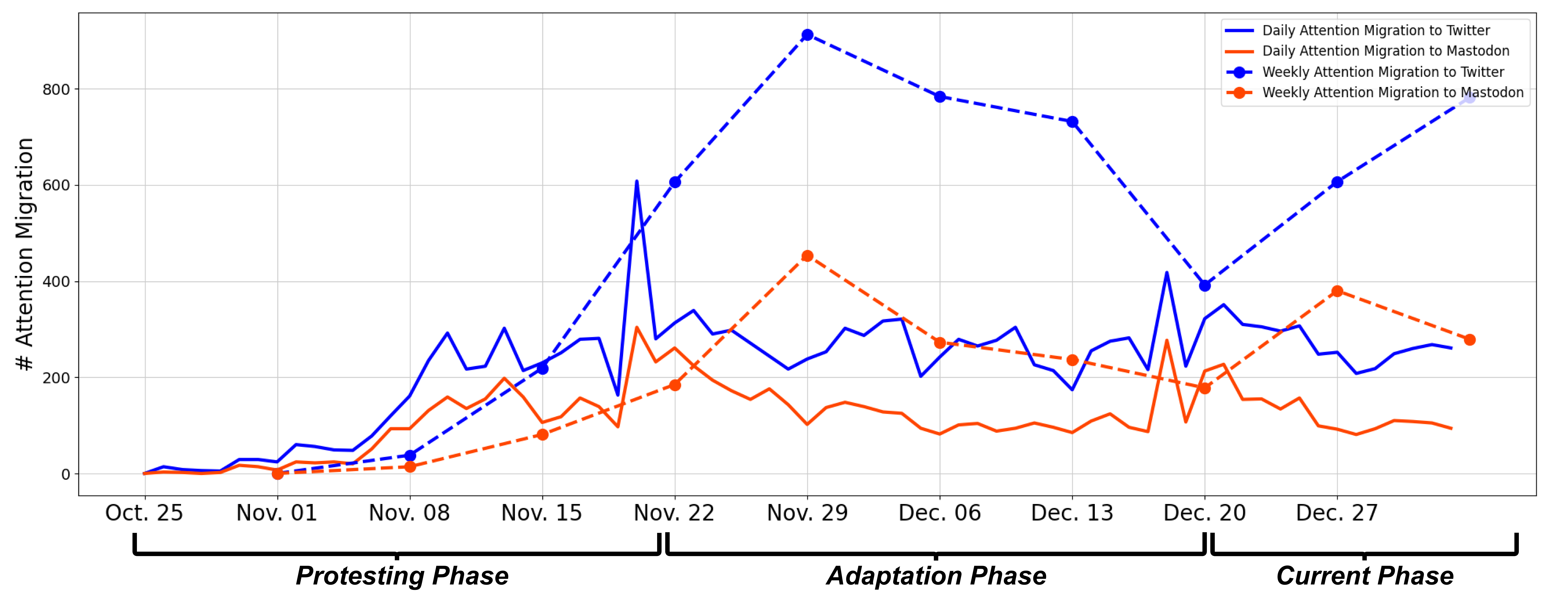}
    
    \caption{Trends in attention migrations toward Twitter (blue) and Mastodon (orange). The $x$-axis represents a date $t_j$ when users migrated their attention. The trends are shown on a daily or weekly basis, with corresponding intervals of $\delta$ = \textit{1 day} or $\delta$ = \textit{1 week}.}
    \label{attention_migration_trend}
\end{figure*}
    

In this section, we investigate the extent to which individuals who were active on Twitter and Mastodon and how they shifted their attention to each platform following Musk's acquisition of Twitter and other major events related to Twitter.

\subsection{Understanding Evolving Migration Patterns}
Figure~\ref{mastodon_active_user_trend} shows the trend between Twitter-related events and the number of daily active users, covering users active on one or both platforms within specified intervals. User activity is assessed over a one-day interval, starting at midnight of the preceding day ($t_i$) and ending at midnight of the current day ($t_j$). For example, setting $t_j$ to January 2nd means the interval covers January 1st from 00:00 to 23:59.

To delineate major changes in Figure~\ref{mastodon_active_user_trend}, we examine temporal shifts in daily active users using Prophet~\cite{taylor2018forecasting} developed by Facebook, a time-series method that is capable of handling nonlinear growth and seasonal changes. Our analysis involved applying Prophet separately to the three trends named Twitter-only, Mastodon-only, and Twitter \& Mastodon. When pinpointing key moments where these temporal shifts overlapped, we found that these overlapped points correspond to Mastodon's first peak (November 22, 2022), and its second peak (December 20, 2022). Based on these two important moments, we were able to distinguish three distinct phases in migration patterns: (1) the \textit{protesting phase}, (2) the \textit{adaptation phase}, and (3) the \textit{current phase}.

The \textit{protesting phase} began following the change in ownership of Twitter and concluded before the first peak in usage of Mastodon. During this phase, the number of active users on Twitter consistently declines, while there is a steep increase in the number of active users on Mastodon. These changes in the number of active users appear with various events happened on the Twitter, such as the mass layoffs at Twitter and the launch of Twitter Blue, which includes the subscription service for verified accounts and policy updates.


The \textit{adaptation phase} follows the first peak and continues until the second peak of Mastodon. In response to recent policy changes on Twitter, there is a notable change in active user engagement on Mastodon. Especially, the number of active users on Mastodon temporarily surges after the relaunch of Twitter Blue and the suspension of Mastodon's account and several high-profile journalists on Twitter~\footnote{https://www.nytimes.com/2022/12/18/business/twitter-ban-social-media-competitors-mastodon.html}. The similar upward trend is noticeable for users active on both platforms.

The \textit{current phase} follows the second peak of Mastodon and continues until the present time. After Elon Musk announced his intention to step down as CEO of Twitter when he finds a suitable successor, the number of active users on Mastodon started to decline. This event may have had a mitigating effect on attention given to Mastodon. At present, it appears that migrants tend to stay active on Twitter rather than continuing to use Mastodon, which raises questions about the longevity of the platform migration.

Figure~\ref{attention_migration_trend} presents the trends of attention migration on daily and weekly intervals. These results indicate that the number of users shifting their attention to Twitter surpasses those moving to Mastodon, especially after each transition point between the migration phases. Notably, migrated users tend to shift more to Twitter after experiencing both platforms. The largest gap in attention migration trends between Twitter and Mastodon emerges during the \textit{adaptation phase}. We speculate such a large gap in attention migration is because numerous users faced challenges in adjusting to Mastodon after arrival. Overall, the findings reveal that migrated users tend to stay on Twitter, and suggest that the momentum of migration to Mastodon has weakened.

\subsection{Inferring Motivational Factors of Migration}
To examine if any events on Twitter prompted users
to migrate to Mastodon, we leveraged BERTopic~\cite{grootendorst2022bertopic} as a tool for examining temporal variations in their discourses. The frequency of tweets and toots relating to the top ten topic groups over time are graphically represented in Figure~\ref{topics_over_time}. Our experiment uncovered a significant portion of Twitter-related discourses (topic group 1 on both Twitter and Mastodon), as well as a strong focus on Elon Musk (topic group 4 on Twitter and topic group 1 on Mastodon). The prominent presence of Twitter-related discourses during the \textit{protesting phase}, coupled with Elon Musk-related discourses toward the end of \textit{adaptation phase}, offer quantitative evidence of a connection between Twitter’s change in ownership and the subsequent platform migration. Furthermore, our analysis revealed that 13.6\% of the total toots included the keywords ``Twitter'' or ``Elon Musk,'' while only 6.1\% of the total tweets contained these keywords. This difference shows that migrated users were proportionally more inclined to discuss Twitter and Elon Musk on Mastodon than on Twitter.

Identifying trending discourses, however, does not provide insight into the motivations for the migration. To fill this gap, we utilized sentiment analysis, leveraging a DeBERTa-based model~\cite{he2020deberta, yang2021back}, to investigate if Musk's acquisition led users to leave Twitter as a form of protest. This model, adeptly fine-tuned for aspect-based sentiment analysis, allowed us to gauge sentiments specifically targeted at either Twitter or Elon Musk, in addition to capturing the general sentiment from a non-specific target.

Figure~\ref{compare_sentiments} illustrates the distribution of sentiment scores of tweets and toots on each platform. Our observations suggest that Mastodon may be particularly appealing to individuals with negative sentiments toward Twitter and its leadership. The migration from Twitter to Mastodon aligns with the push-pull theory of migration~\cite{lee1966theory}, wherein the relatively negative sentiment toward Twitter and its leadership may act as a ``push factor"~\cite{newell2016user} for users to leave Twitter. In contrast, users show more positive sentiment when Mastodon is a target than Twitter, which can serve as a ``pull factor'' attracting users to Mastodon~\cite{hou2020understanding}.

\begin{figure*}
    \centering
    \includegraphics[width=1.0\textwidth]{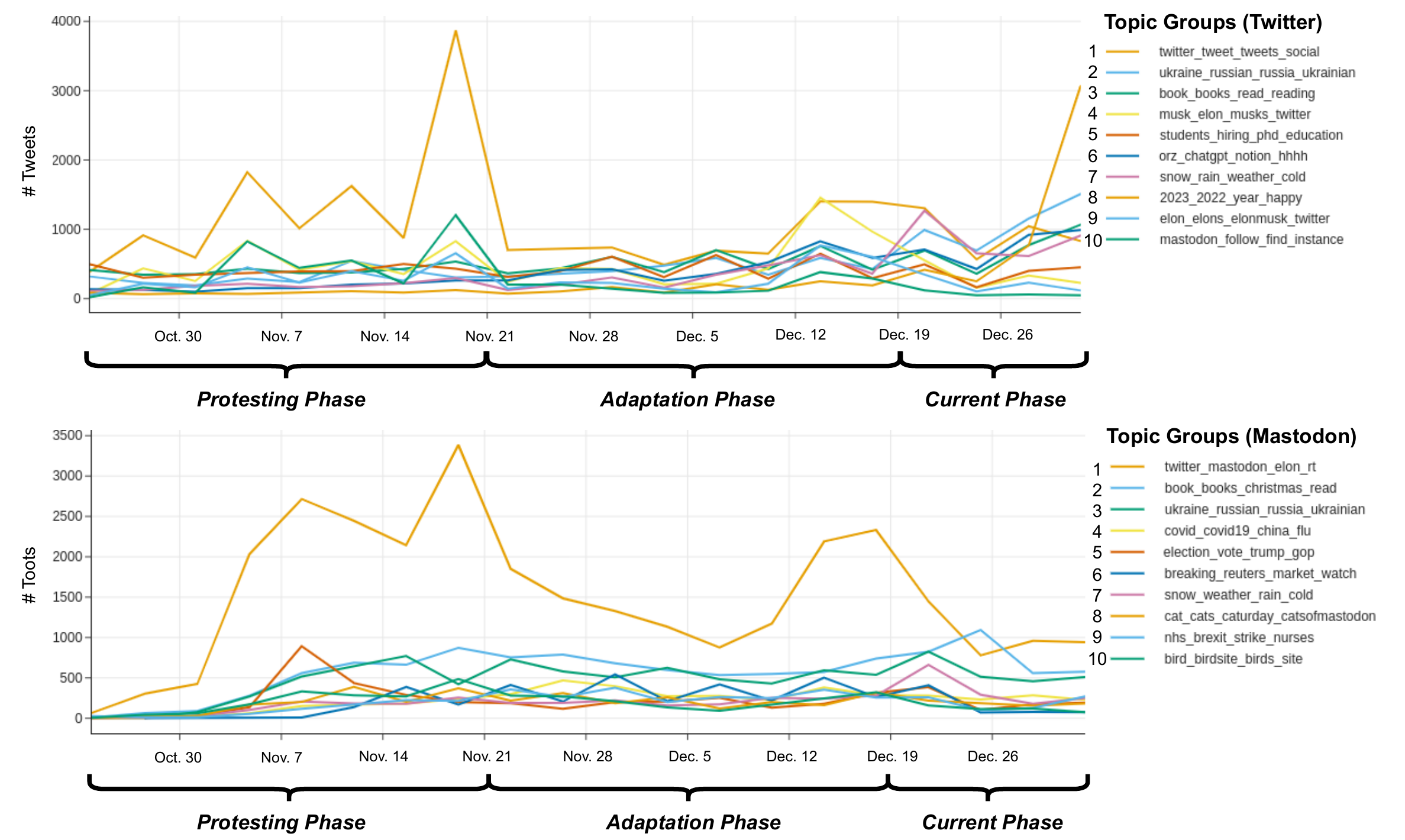}
    
    \caption{The trends in the number of tweets on Twitter and toots on Mastodon that mention one of the top 10 topic groups. Each topic group is shown with a discourse consisting of its four most frequently occurring words, separated by  underscores. The topic groups are numbered based on their rank of proportion to the total messages, where a lower rank signifies a higher proportion. Note that the $x$-axis represents the exact date when the tweet or toot was published. As a result, the dates for the protesting, adaptation, and current phases are a day earlier than those referenced in Figure~\ref{mastodon_active_user_trend}.}
    \label{topics_over_time}
\end{figure*}

\begin{figure*}
    \centering
    \includegraphics[width=0.98\textwidth]{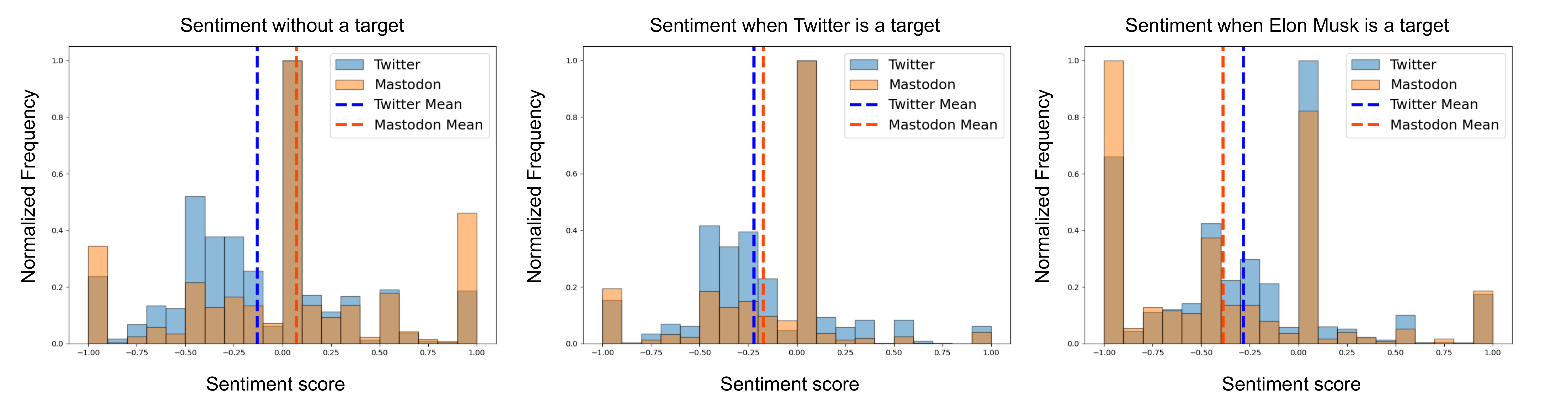}
    \caption{The sentiment score distribution is compared between Twitter (blue) and Mastodon (orange). We compare the distributions of Twitter and Mastodon across three categories: (1) sentiment without a target; (2) sentiment with Twitter as a target ; and (3) sentiment with Elon Musk as a target. The scores range from -1 to 1, with lower values indicating a more negative sentiment. The mean score for each distribution is shown by the dashed lines. The Kolmogorov-Smirnov test revealed significant sentiment differences between Twitter and Mastodon in all three categories, with a significant $p$-value of less than 0.001.}
    \label{compare_sentiments}
\end{figure*}

\section{RQ2: Distinct User Behaviors of Migrated Users on Twitter and Mastodon}
Addressing this research question allows us to compare how migrated users exhibit their behaviors distinctively on Twitter and Mastodon, and discern if their varying behaviors can be explained by the unique platform architectures. Moreover, we also aim to uncover the disparities between these platforms based on the responses that users attracted on each platform.


\subsection{User-level Features}

To explore how users exhibit different behaviors on each platform, we decided to compare the common user behaviors that are shared between Mastodon and Twitter, such as tweets and toots. Given a specific user $u$ and a platform $p$, we define three distinct types of user-level features on each platform: (1) \textit{User Activity}, (2) \textit{User Network}, and (3) \textit{User Response}. The definitions for these categories are as follows:


\paragraph{\textbf{User Activity}.} $\mathcal{A}(u, p)$ is determined by a user's two activity types denoted as $\{a_i\}^2_{i=1}$, which compromise status messages and reshares (e.g., tweets and retweets a user has posted). The counting function $\sigma(u, a_i, p)$ measures the cumulative count of the activity $a_i$ during the studied period. Formally, the user activity can be represented as:

\begin{equation}
\mathcal{A}(u, p) = \sum\limits_{i=1}^{2} \frac{\sigma(u, a_i, p) - \min_{u'}\left(\sigma(u', a_i, p)\right)}{\max_{u'}\left(\sigma(u', a_i, p)\right) - \min_{u'}\left(\sigma(u', a_i, p)\right)}
\end{equation}

\paragraph{\textbf{User Network}.} $\mathcal{W}(u, p)$ is derived by two network types denoted as $\{w_i\}^2_{i=1}$, which include followers and followees of a user. The counting function $\sigma(u, w_i, p)$ measures the size of the network $w_i$. The user network can be defined as:

\begin{equation}
\mathcal{W}(u, p) = \sum\limits_{i=1}^{2}\frac{\sigma(u, w_i, p) - \min_{u'}\left(\sigma(u', w_i, p)\right)}{\max_{u'}\left(\sigma(u', w_i, p)\right) - \min_{u'}\left(\sigma(u', w_i, p)\right)}
\end{equation}

\paragraph{\textbf{User Response}.} $\mathcal{R}(u, p)$ is obtained by two response types denoted as $\{r_i\}^2_{i=1}$, which include favorites and reshare a user has received (e.g., likes and retweets a user received). The counting function $\sigma(u, r_i, p)$ measures the cumulative count of the response $r_i$ during the studied period. Formally, the user response can be represented as:

\begin{equation}
\mathcal{R}(u, p) = \sum\limits_{i=1}^{2} \frac{\sigma(u, r_i, p) - \min_{u'}\left(\sigma(u', r_i, p)\right)}{\max_{u'}\left(\sigma(u', r_i, p)\right) - \min_{u'}\left(\sigma(u', r_i, p)\right)}
\end{equation}
where $u'$ denotes a user within the population under study. Please note that we utilize max-min normalization on each platform when calculating $\mathcal{A}(u, p)$, $\mathcal{W}(u, p)$, and $\mathcal{R}(u, p)$. This normalization accounts for the different scales of user behaviors present on each platform.


\subsection{Disparity in User-level Features}
\label{estimating}

We conducted a comparative analysis of three distinct user-level features on both Twitter and Mastodon platforms. Our comprehensive study derives insights from Table~\ref{user_features_comparison_table}, which shows the mean and mean absolute deviation values, along with Figure~\ref{user_features_comparison_figure}, that illustrates the interquartile plot. Based on this analysis, we make three primary observations as follows:

\begin{figure}
    \centering
    \includegraphics[width=0.40\textwidth]{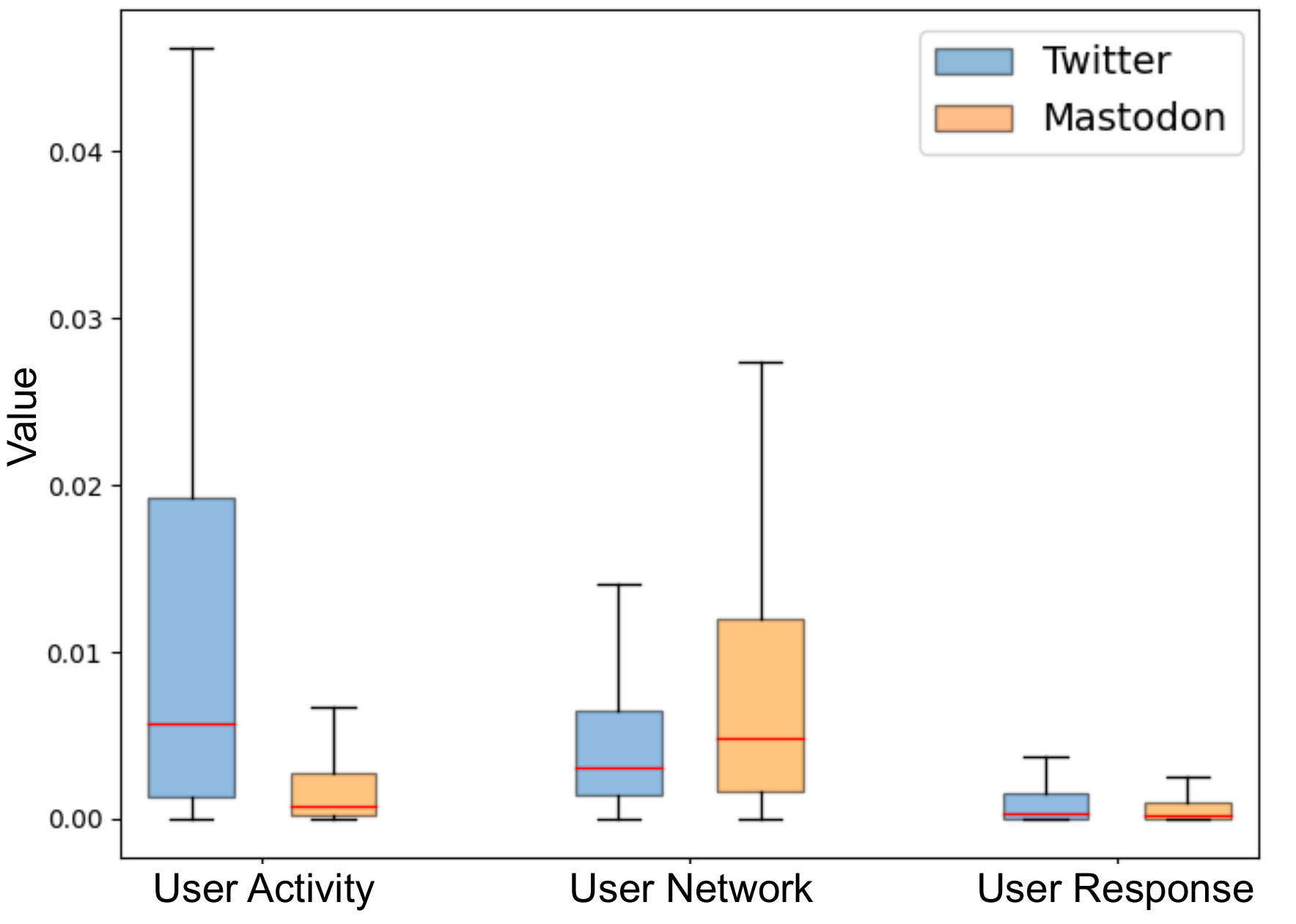}
    \caption{Interquartile box plots comparing the values of user-level features in Twitter (blue) and Mastodon (orange). Within each box plot, the red line indicates the median value.}
    \label{user_features_comparison_figure}
\end{figure}

First, we observe that users exhibit, on average, $\sim$4 times higher \textit{user activity} on Twitter than on Mastodon. This disparity may stem from the fact that numerous users continued to use both platforms even after migration, with Twitter serving as their primary platform~\cite{he2023flocking}. Additionally, the higher MAD value on Twitter indicates a broader spread of activities compared to Mastodon.

Second, we find that users show, on average, $\sim$1.4 times more \textit{user network} on Mastodon than on Twitter. This indicates that users tend to focus on building connections with other users on Mastodon~\cite{shaw2020decentralized} than Twitter. The MAD value is higher on Mastodon, signifying a more varied distribution of network sizes among the migrated users.

Third, our analysis reveal that users, on average, receive $\sim$1.3 times more \textit{user response} on Twitter than on Mastodon, likely due to Mastodon's lower Q1 value (0.00003) compared to Twitter's (0.003). The higher MAD value on Twitter suggests that user responses are more well spread among users on Twitter. Notably, a larger percentage of users receive zero responses on Mastodon (13.6\%) than on Twitter (3.2\%), consistent with previous findings of overall lower response rates on Mastodon~\cite{la2022network}.

\begin{table}
\centering
\small 
\begin{tabular}{ccccccc}
\toprule
\multicolumn{1}{c}{\multirow{2}{*}{\textbf{Metric}}}
& \multicolumn{2}{c}{\textbf{$\mathcal{A}$}}
& \multicolumn{2}{c}{\textbf{$\mathcal{W}$}}
& \multicolumn{2}{c}{\textbf{$\mathcal{R}$}} \\
\cmidrule(l){2-3} \cmidrule(l){4-5} \cmidrule(l){6-7}
& A & B & A & B & A & B \\
\midrule
AVG & 0.020 & 0.005 & 0.007 & 0.010 & 0.005 & 0.004\\
MAD & 0.023 & 0.006 & 0.006 & 0.010 & 0.008 & 0.006\\
\bottomrule
\end{tabular}
\caption{Statistics of user-level features in platform \textbf{A} (Twitter) and \textbf{B} (Mastodon). AVG is average, and MAD (Mean Absolute Deviation) shows the dispersion from the average.}
\label{user_features_comparison_table}
\end{table}

\subsection{Disparity in Occupations to User Response}

\begin{figure}
    \centering
    \includegraphics[width=0.44\textwidth]{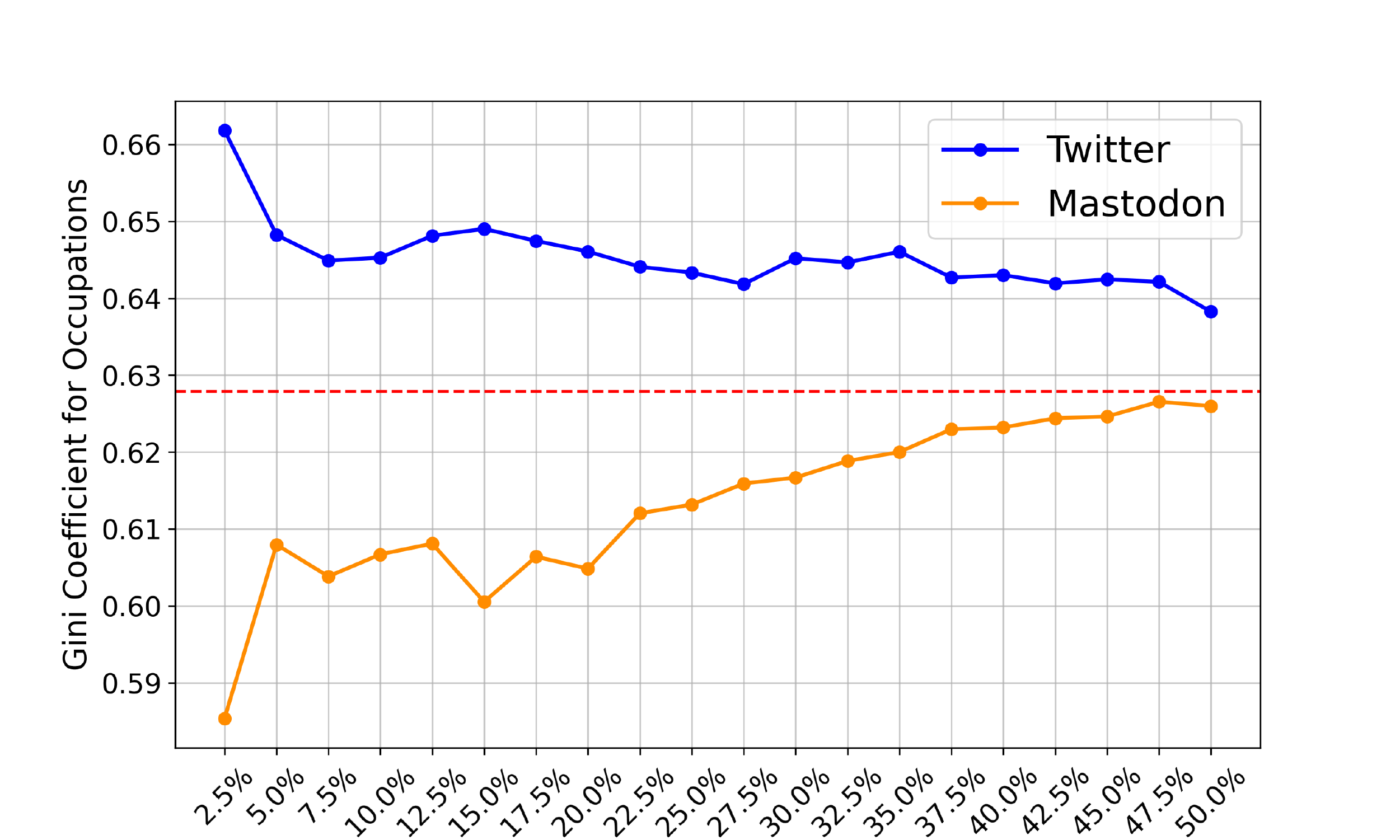}
    \caption{Gini coefficients for Twitter (blue) and Mastodon (orange) depict occupational distribution among migrants, focusing on users above the top 50\% response rate. The red dashed line is the Gini coefficient for the original sample.}
    \label{gini_percentile_trend}
\end{figure}

We utilized the Gini coefficient, a widely accepted metric for assessing disparities in income or wealth among diverse groups~\cite{gini1936measure}, to evaluate the inequality in occupational distribution. The Gini coefficient values range between 0 and 1, with lower values signifying greater equality. We computed the Gini coefficient using $n$ occupation groups (i.e., the nine major groups in the UK's SOC code) and $x_i$, representing the user count for the $i$-th group among the 5,817 users (58.1\% of the 10,011 users) with identifiable occupations as follows:

\begin{equation}
Gini\:Coefficient = 1 - \sum_{i=1}^{n} \left(\frac{x_i}{\sum_{j=1}^{n} x_j}\right)^2 \end{equation}

The Gini coefficient for 5,817 users is calculated to be 0.627. Such high inequality mainly stems from the overrepresentation of user groups like SOC 2xx (Professionals) and SOC 3xx (Associate Professionals and Technicians). One potential reason for this inequality is the unique appeal that Mastodon's decentralized platform holds for individuals with academic or technology-related backgrounds~\cite{kupferschmidt2022musk,shaw2020decentralized,zignani2018follow}.


To discern if certain groups of occupations are encouraged to attract more user responses on each platform, we examined the occupational inequality of a subset in each platform. Our primary focus was on the relationship between occupational inequality and the user response on each platform. To this end, we first ranked users according to their responses on each platform. Once we created the ranking, we focused on the users who actively engage on each platform and these users represent a value above the median in the data.

Figure~\ref{gini_percentile_trend} demonstrates that occupational inequality on Twitter surpasses the inequality of the original sample within the top 50\% of users. Interestingly, the disparity between Twitter and Mastodon peaks within the highest 2.5\% of users, underlining a significant inequality among this extremely popular user group on Twitter. Conversely, on Mastodon, users exhibit lower inequality than the original sample up to the top 50\% of users, suggesting a more evenly distributed occupational distribution among its highly engaged users.

\subsection{Disparity in Hashtags to User Response}

We compared the unique set of hashtags shared by each user with their popularity on Twitter and Mastodon. Given that a user's popularity on one platform, gauged by the size of their user response, may not necessarily translate to popularity on another platform, we grouped users into four categories based on their popularity: (1) popular on Twitter, but not Mastodon; (2) popular on Mastodon, but not Twitter; (3) popular on both platforms; and (4) unpopular on both platforms.

Users were grouped into one of four categories based on their ranks on each platform, $rank_A$ for Twitter and $rank_B$ for Mastodon. For categories (1) and (2), users are sorted in descending order based on the difference in their ranks ($rank_{diff} = rank_A - rank_B$). This represents users with the largest differences in user response between the two platforms, with the top 10\% being users who are popular on Mastodon but not on Twitter and the bottom 10\% being users who are popular on Twitter but not on Mastodon. For categories (3) and (4), users are sorted based on the sum of their ranks ($rank_{sum} = rank_A + rank_B$). The top 10\% in this case are users who are unpopular on both platforms, while the bottom 10\% were users who are popular on both platforms.

\begin{figure*}
    \centering
\includegraphics[width=0.98\textwidth]{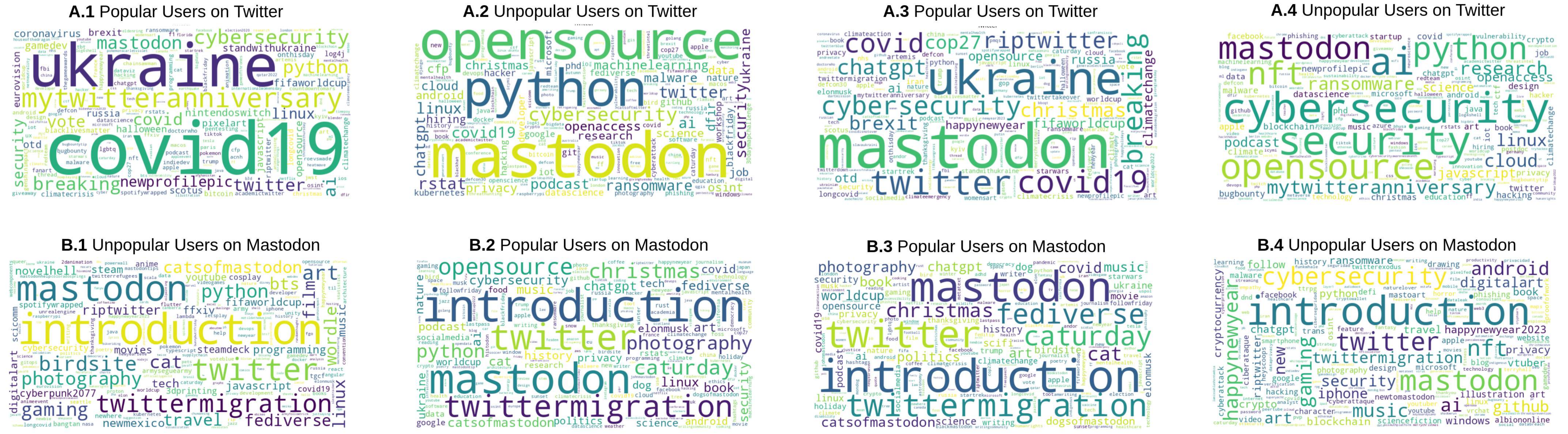}
    \caption{Wordclouds of unique lowercase hashtags on two platforms \textbf{A} (Twitter) and \textbf{B} (Mastodon). Hashtags are categorized based on the popularity of migrants: (1) popular on Twitter but not on Mastodon, (2) popular on Mastodon but not on Twitter, (3) popular on both platforms, and (4) unpopular on both platforms. More frequent hashtags are highlighted with larger font size.}
    \label{hashtags_user_response}
\end{figure*}

Figure~\ref{hashtags_user_response} showcases frequently used unique hashtags, classified into four categories based on the popularity of users on each platform. On Twitter, we observed popular users predominantly focus on social issue hashtags, such as \texttt{\#Covid} and \texttt{\#Ukraine}, whereas their less popular counterparts are inclined toward technology-oriented discussions, evident by those frequent hashtags such as \texttt{\#Opensource}, \texttt{\#Security}, and \texttt{\#Python}. This suggests sharing global and social hashtags tend to be the main focus for getting more response on Twitter. In contrast, Mastodon users of all popularity levels engage with niche contents, encompassing photography, gaming, cats, and programming. Notably, users often share migration-specific hashtags on Mastodon, such as \texttt{\#TwitterMigration} and \texttt{\#Introduction}, which are relatively uncommon among the  users popular on Twitter.
\subsection{Analysis Summary}
Our findings suggest empirical evidence indicating that the unique architectures of these platforms influence two aspects:

\begin{itemize}[leftmargin=*]
\item  The occupational inequality among popular users is more pronounced on Twitter than on Mastodon. This can be due to Twitter's design, which focuses on popularity by promoting well-known users and their posts~\cite{la2021understanding}. In contrast, Mastodon's topic-oriented approach creates connections and interactions based on shared interests. This approach provides an opportunity for migrated users from diverse occupations to gain popularity.
\item Popular users on Twitter often share hashtags centered on global and societal issues. We speculate that this tendency could be linked to Twitter's central structure, which enables broader reach and a larger audience, thereby amplifying the popularity of users with social significance~\cite{kwak2010twitter, lehmann2012dynamical}. Conversely, Mastodon users create connections by seeking out those with similar interests, promoting the sharing of more niche content, relatively irrespective of the users' popularity levels.
\end{itemize}

\noindent Future research aims to enhance our understanding of the communication dynamics between Twitter and Mastodon. To investigate how information consumption is influenced among migrating users, further studies should focus on: (1) the impact of disparities between Twitter and Mastodon on the communication structure of both platforms; (2) the effects of disparities in occupational distribution on the overall user experience and information sharing; and (3) how these imbalances can inform strategies to improve user experience and promote equitable engagement among diverse users.




\section{RQ3: Sustainability of Platform Migration: \\A Case Study of Mastodon}
\subsection{Behavioral Characteristics of Residents}


In our investigation of user residency on Mastodon, we categorized the migrated users into two distinct groups: (1) \textit{Non-residents}, whose last activity on Mastodon was either during the protesting or adaptation phase, and (2) \textit{Residents}, whose last activity on Mastodon is during the current phase.


In analyzing the behavioral aspects that retain users on Mastodon, we applied user-level features to predict the two types of migrants. This investigation involved unique user-level features for Mastodon such as (1) Interaction Diversity, (2) Fandom Migration, and (3) Migration Hashtags.

\paragraph{\textbf{Interaction Diversity.}} We conducted a study on interaction diversity for a specific user, denoted as $u$. This metric measures the level of interaction for a user $u$ to engage with
other users from different primary servers. The Shannon entropy formula is utilized to compute this metric as follows:
\begin{equation}
\begin{aligned}
H(u) = - \sum_{i=1}^{|\mathcal{S}|} P(u, s_i) \log P(u, s_i),\\
P(u, s_i) = \frac{\sigma(u, s_i)}{\sum_{j=1}^{|\mathcal{S}|} \sigma(u, s_i)},
\end{aligned}
\end{equation}

\noindent where $\sigma(u, s_i)$ is the cumulative count of interactions with users from server $s_i \in \mathcal{S}$, with interactions defined as receiving follows or replies by other users. $P(u, s_i)$ is the probability of user $u$ interacting with users from server $s_i$.


\paragraph{\textbf{Fandom Migration.}} We examined the concept of ``Fandom migration"~\cite{fiesler2020moving}, a phenomenon in which a leader's migration to a new platform often leads to a collective migration of their followers from the previous platform as well. To determine the percentage of a user's Mastodon followers who had previously followed them on Twitter, we analyzed identical usernames on both platforms, under the premise that identical usernames belong to the same person.

\paragraph{\textbf{Migration Hashtags.}} To understand if users who led the migration from Twitter to Mastodon  were more likely to remain, we analyzed the frequency of migration-specific hashtags, including: \texttt{\#RipTwitter, \#GoodbyeTwitter, \#JoinMastodon, \#MastodonMigration, \#MastodonSocial, \#ByebyeTwitter, \#TwitterMigration, \#TwitterTakeover, \#TwitterShutdown, \#LeaveTwitter, \#TwitterRefugee and \#TwitterExodus.}


For our analysis to classify the two types of migrants, namely \textit{non-residents} and \textit{residents}, we performed statistical tests on six distinct user-level characteristics. We found that a total of 5,590 \textit{residents} (55.83\% of the 10,011 users) were present at the time of the study. The outcomes of the logistic regression, after standardization of the feature values, are presented in Table~\ref{statistics_test_user_category}. These results demonstrate the importance of various user-level features in predicting the types of migrants. Among the examined features, user activity, user network size, diversity of interactions, and fandom migration emerge as highly significant features, each demonstrating \textit{p}-values lower than 0.001. Positive coefficients of these features reveal a positive association with the types of migrants, suggesting that as the value of these feature increases, so does the likelihood of migrated users being \textit{residents}. Remarkably, user activity and interaction diversity show distinctively higher coefficients than other features, indicating their strong association with user retention to Mastodon.

Surprisingly, the volume of user response is not significantly associated with a user's decision to stay on Mastodon, implying that users who merely focus on attracting a large user response tend to discontinue using the platform over time. Moreover, there is no statistical correlation between the frequency of sharing migration-specific hashtags and migrant types, indicating that individuals who once actively led the migration movement may not necessarily stay on Mastodon.

\begin{table}
\centering
\small
\begin{tabular}{ccccc}
\toprule
\textbf{Features} & \textbf{Coef.} & \textbf{SE}&\textbf{$p$-value}&\textbf{OR} \\
\midrule
User Activity & 0.914 & 0.081 & 0.000** &2.494 \\
\midrule
User Network & 0.173 & 0.040 & 0.000** &1.189 \\
\midrule
User Response & -0.053 & 0.032  & 0.100 &0.948 \\
\midrule
Interaction Diversity & 0.724 & 0.027 & 0.000** &1.077 \\
\midrule
Fandom Migration & 0.074 & 0.022 & 0.000** & 1.076 \\
\midrule
Migration Hashtags & -0.041 & 0.028 & 0.147 & 0.959 \\
\bottomrule
\end{tabular}
\caption{Results on logistic regression to predict the two types
of migrants (\textit{non-resident} and \textit{resident}) on Mastodon. Coef represents coefficient, SE stands for standard error, and OR denotes odd ratio. Note that *$p<0.05$ and **$p<0.001$. }
\label{statistics_test_user_category}
\end{table}

\subsection{Analysis Summary}
Mastodon, a decentralized social media platform, uses a federated network to encourage connections based on shared interests. However, this can lead to social fragmentation, complicating user interaction during building social networks.~\cite{raman2019challenges}. Our analysis yields two notable outcomes:

\begin{itemize}[leftmargin=*]
\item The decision to remain on Mastodon primarily is associated with a user's active effort to frequent conversations and diverse social interactions to appreciate the community-centered experience, rather than by the volume of responses received or the frequent sharing of migration hashtags.
\item Mastodon distinguishes itself by prioritizing community-centered experiences, contrasting with traditional platforms that often give preference to individual self-promotion or the creation of viral content to capture a broader audience.
\end{itemize}

\noindent As a prospective avenue to enhance the sustainability of platform migration, it may be helpful to examine social network analysis theories that can shed light on concealed patterns within the social networks of migrated users by pinpointing information brokers linking disconnected groups, or structural holes~\cite{burt2018structural} and by examining the impact of weak ties in loosely connected networks on obtaining unique information and resources~\cite{granovetter1973strength}. In turn, this information can inform platform developers, community managers, and users about the factors that facilitate sustainability of migration to a platform with distinct architecture and reduce mass departures and provide insight on how to avoid similar issues when launching new social media.

\section{Limitations}

First, the data available for this study do not include the 266 users who experienced \textit{permanent migration} by January 2, 2023, due to API restrictions preventing access to deleted accounts on both Twitter and Mastodon. Our data is also limited to the first ten weeks post Elon Musk's Twitter acquisition. As such, the findings from the migration patterns might not be applicable to other periods or predictive of future trends.




Second, we acknowledge the presence of alternative platforms for Twitter, such as Hive (a microblogging service) and Damus (decentralized network powered by Nostr protocol). However, our study focuses on the migration from Twitter to Mastodon, assuming that user attention mainly oscillates between Twitter and Mastodon due to Mastodon's increased popularity during the time that Musk acquired Twitter.

Last, our study does not provide a causal analysis. Instead, the motivational factors are inferred from observed data on migration patterns, discourses over time, and users' sentiment towards them. Testing the correctness of the inferred motivations for migration requires surveys of users including open-ended interviews, where users can report their
reasons for migrating or not from any platform.

\section{Conclusion}
We introduce two types of migration on social media and explore the migration patterns between Twitter and Mastodon. Our observations indicate that dissatisfaction with Twitter and its management, notably the Twitter's ownership change, is a primary motivation for users to migrate to Mastodon. Interestingly, these migrated users often utilize both platforms, frequently shifting their attention back to Twitter after trying both platforms. This migration pattern refutes claims of dooms day scenario for Twitter after the mass exodus.


Our comparison of user behaviors on Twitter and Mastodon unveils notable disparities between the two platforms. Specifically, users exhibit a larger inequality in the distribution of users' occupations within highly engaged users on Twitter. An analysis of hashtag usage reveals that popular Twitter users typically concentrate on global and social issues, while Mastodon users, regardless of their popularity, often share hashtags related to migration and niche content.

We identify key user behaviors that are associated with the retention of users on Mastodon, with the variety of interactions with users across different Mastodon servers being a highly significant factor. Conversely, the volume of user responses, such as favorites and reshares, does not exhibit a significant impact on user retention. This underscores the distinctive appeal of Mastodon's features, such as its decentralized architecture and community-oriented interactions, and how they relate to user behaviors.

In conclusion, our findings offer valuable insights for platform migration and how to retain users during periods after a mass migration. By leveraging these behavioral factors, platform designers can create more engaging and sustainable platforms that cater to user preferences and needs.





\section{Future Work}
Exploring the phenomenon of returning migration is an important area for future research. Our plan is to study this phenomenon as part of a long-term analysis when more user migration data becomes available between Twitter and Mastodon~\cite{jeong2023user}. Furthermore, we aim to understand how the disparities between Twitter and Mastodon impact each platform, including user experience and the way information is shared among migrated users, including the spread of disinformation~\cite{jeong2022classifying, jeong2022nothing}. One aspect of this analysis is to apply theories of weak ties~\cite{granovetter1973strength} and structural holes~\cite{burt2018structural}. Furthermore, surveys will be conducted to complement the findings and provide qualitative evidence of the potential motivations behind migration. Finally, the proposed approach will be extended to other alternative social media platforms beyond Twitter, to gather more comprehensive understanding of the migration. 

\section{Appendix}

\subsection{Standard Occupational Classification (SOC)}
The UK's SOC 2010 system uses a hierarchical structure to organize occupations, where the first digit of each SOC code represents the nine major groups, as shown in  Table~\ref{tab:uk_soc_categories}

\begin{table}[h]
\centering
\small
\begin{tabular}{|>{\centering\arraybackslash}m{1.5cm}|>{\centering\arraybackslash}m{6cm}|}
\hline
\textbf{Major Group} & \textbf{ Description} \\
\hline
1xx & Planing, directing, or coordinating the operations of businesses or other organizations. \\
\hline
2xx & Specialized tasks that require advanced knowledge, such as law, science, or education. \\
\hline
3xx & Support in various professional fields, such as healthcare, science, or engineering. \\
\hline
4xx & Administrative and clerical duties for business, government agencies or organizations. \\
\hline
5xx & Work in skilled jobs, such as construction, mechanics, or electrical work. \\
\hline
6xx & Personal services to individuals or groups, such as childcare, leisure, or personal care. \\
\hline
7xx & Selling products or services to customers or provide customer support. \\
\hline
8xx & Operating machinery or perform manual labor in manufacturing or production. \\
\hline
9xx & Simple and routine tasks that do not usually require formal education or training. \\
\hline
\end{tabular}
\caption{Nine major groups in the UK's SOC 2010 system.}
\label{tab:uk_soc_categories}
\end{table}

\subsection{Experimental Details}
Utilizing the Prophet model set to change point prior scale of 0.252 and change point range of 0.95 with linear growth, we observed eight temporal shifts in Twitter-only users on specific dates (Nov. 1, 9, 22, 24, Dec. 5, 10, 20, 28), five in Mastodon-only users (Nov. 11, 19, 22, Dec. 7, 20), and five in Mastodon \& Twitter users (Nov. 11, 19, 22, Dec. 5, 20). In BERTopic model, we utilized \texttt{all-MiniLM-L6-v2} as the default model, which is a pre-trained transformer model designed for embedding English sentences and short paragraphs. Focusing on toots and tweets in English, we optimized our system settings to auto-adjust with a minimum of 100 topics, which resulted in generating 841 topics for tweets and 483 for toots. For aspect-based sentiment analysis, we leveraged the DeBERTa with \texttt{deberta-v3-large-absa}, which is a fine-tuned model for ABSA (Aspect-Based Sentiment Analysis) datasets. The model's default parameter setup has a hidden size of 768, 12 hidden layers, and 12 attention heads.


\section{Ethical Statement, Impact, and Reproducibility}
\label{reproducibility}
Our proposed approach for analyzing migration from Twitter to Mastodon may suffer from selection bias, as it relies on users who have chosen to disclose their other social media platform accounts on their profiles. As a result, the selected users may not be representative of the larger population of users migrating from Twitter to Mastodon.


Our study aims to gain insights into the migration of users between Twitter and Mastodon, and is not intended to exploit or manipulate these communities for financial gain. Our data collection was completed before Twitter updated its basic API tier to a priced model. In compliance with the terms of service of Twitter and Mastodon, we disclose only the user IDs of the accounts included in our study. The codes used for mapping user accounts between Twitter and Mastodon are available at our repository\footnote{https://github.com/ujeong1/ICWSM24-Exploring-Platform-Migration-Patterns-between-Twitter-and-Mastodon}

\section{Acknowledgments}
This work received support from the Office of Naval Research (Award No. N00014-21-1-4002) and the National Science Foundation (Award No. 2227488). Opinions, interpretations, conclusions, and recommendations within this article are solely those of the authors. Special acknowledgment goes to Dr. Kaize Ding, Assistant Professor at Northwestern University, whose insightful feedback was instrumental in improving the quality of the article.


\fontsize{9.0pt}{10.0pt} \selectfont
\bibliography{main}

\appendix

\subsection{Paper Checklist}

\begin{enumerate}

\item For most authors...
\begin{enumerate}
    \item  Would answering this research question advance science without violating social contracts, such as violating privacy norms, perpetuating unfair profiling, exacerbating the socio-economic divide, or implying disrespect to societies or cultures?
    \answerYes{Yes, see the Ethical Statement, Impact, and Reproducibility.}
  \item Do your main claims in the abstract and introduction accurately reflect the paper's contributions and scope?
    \answerYes{Yes, the abstract provides three research questions, and the questions are mentioned and solved in each section respectively.}
   \item Do you clarify how the proposed methodological approach is appropriate for the claims made? 
    \answerYes{Yes, justification is provided for each methodology when answering research questions.}
   \item Do you clarify what possible artifacts in the data used, given population-specific distributions?
    \answerYes{Yes, see the Limitations.}
  \item Did you describe the limitations of your work?
    \answerYes{Yes, see the Limitations.}
  \item Did you discuss any potential negative societal impacts of your work?
    \answerYes{Yes, see the Ethical Statement, Impact, and Reproducibility.}
      \item Did you discuss any potential misuse of your work?
    \answerYes{Yes, see the Ethical Statement, Impact, and Reproducibility.}
    \item Did you describe steps taken to prevent or mitigate potential negative outcomes of the research, such as data and model documentation, data anonymization, responsible release, access control, and the reproducibility of findings?
    \answerYes{Yes, see the Ethical Statement, Impact, and Reproducibility.}
  \item Have you read the ethics review guidelines and ensured that your paper conforms to them?
    \answerYes{Yes, see the Ethical Statement, Impact, and Reproducibility.}
\end{enumerate}

\item Additionally, if your study involves hypotheses testing...
\begin{enumerate}
  \item Did you clearly state the assumptions underlying all theoretical results?
    \answerNA{NA}
  \item Have you provided justifications for all theoretical results?
    \answerNA{NA}
  \item Did you discuss competing hypotheses or theories that might challenge or complement your theoretical results?
    \answerNA{NA}
  \item Have you considered alternative mechanisms or explanations that might account for the same outcomes observed in your study?
    \answerNA{NA}
  \item Did you address potential biases or limitations in your theoretical framework?
    \answerNA{NA}
  \item Have you related your theoretical results to the existing literature in social science?
    \answerNA{NA}
  \item Did you discuss the implications of your theoretical results for policy, practice, or further research in the social science domain?
    \answerNA{NA}
\end{enumerate}

\item Additionally, if you are including theoretical proofs...
\begin{enumerate}
  \item Did you state the full set of assumptions of all theoretical results?
    \answerNA{NA}
	\item Did you include complete proofs of all theoretical results?
    \answerNA{NA}
\end{enumerate}

\item Additionally, if you ran machine learning experiments...
\begin{enumerate}
  \item Did you include the code, data, and instructions needed to reproduce the main experimental results (either in the supplemental material or as a URL)?
    \answerYes{Yes, see the Ethical Statement, Impact, and Reproducibility.}
  \item Did you specify all the training details (e.g., data splits, hyperparameters, how they were chosen)?
    \answerNA{NA}
     \item Did you report error bars (e.g., with respect to the random seed after running experiments multiple times)?
    \answerNA{NA}
	\item Did you include the total amount of compute and the type of resources used (e.g., type of GPUs, internal cluster, or cloud provider)?
    \answerNA{NA}
     \item Do you justify how the proposed evaluation is sufficient and appropriate to the claims made? 
    \answerYes{Yes, see the Data Collection.}
     \item Do you discuss what is ``the cost`` of misclassification and fault (in)tolerance?
    \answerNA{NA}
  
\end{enumerate}

\item Additionally, if you are using existing assets (e.g., code, data, models) or curating/releasing new assets, \textbf{without compromising anonymity}...
\begin{enumerate}
  \item If your work uses existing assets, did you cite the creators?
    \answerYes{Yes, see the Data Collection and Research Questions (RQ1 and RQ2) for the assets we used for the reseasrch.}
  \item Did you mention the license of the assets?
    \answerNo{No, but we confirmed BERTopic, DeBERTa, StanfordCoreNLP, and Prophet are available under the MIT License. The occupation encoder used in this paper utilizes the UK's SOC system and the code is under the GNU General Public License.}
  \item Did you include any new assets in the supplemental material or as a URL?
    \answerYes{Yes, see the Ethical Statement, Impact, and Reproducibility.}
  \item Did you discuss whether and how consent was obtained from people whose data you're using/curating?
    \answerNA{NA}
  \item Did you discuss whether the data you are using/curating contains personally identifiable information or offensive content?
    \answerYes{Yes, see the Ethical Statement, Impact, and Reproducibility.}
\item If you are curating or releasing new datasets, did you discuss how you intend to make your datasets FAIR?
\answerYes{No, because  Twitter has changed its API to a pricing model, it limits accessibility to the protocol at the moment.}
\item If you are curating or releasing new datasets, did you create a Datasheet for the Dataset? 
\answerNo{No, it is not available at this time due to the terms and conditions on Twitter (or X) for developers. See the Ethical Statement, Impact, and Reproducibility sections for more details.}
\end{enumerate}

\item Additionally, if you used crowdsourcing or conducted research with human subjects, \textbf{without compromising anonymity}...
\begin{enumerate}
  \item Did you include the full text of instructions given to participants and screenshots?
    \answerNA{NA}
  \item Did you describe any potential participant risks, with mentions of Institutional Review Board (IRB) approvals?
    \answerNA{NA}
  \item Did you include the estimated hourly wage paid to participants and the total amount spent on participant compensation?
    \answerNA{NA}
   \item Did you discuss how data is stored, shared, and deidentified?
   \answerNA{NA}
\end{enumerate}

\end{enumerate}
\end{document}